\documentclass[reprint,showpacs,preprintnumbers,amsmath,amssymb,longbibliography,aip]{revtex4-1}
\usepackage[utf8]{inputenc}
\usepackage{mathrsfs}
\usepackage{amsmath}
\usepackage{bm}
\usepackage{amsfonts}
\usepackage{amssymb}
\usepackage{amsthm}
\usepackage{color}
\usepackage{graphicx}
\graphicspath{{./images/}}
\usepackage{geometry}
\usepackage{hyperref}
\usepackage{ulem}
\usepackage{epstopdf}
\usepackage{graphicx}
\hypersetup{
     unicode=false,
     pdftoolbar=true,
     pdfmenubar=true,
     pdffitwindow=false,     
     pdfstartview={FitH},    
     pdftitle={My title},    
     pdfauthor={Author},     
     pdfsubject={Subject},   
     pdfcreator={Creator},   
     pdfproducer={Producer}, 
     pdfkeywords={keyword1} {key2} {key3}, 
     pdfnewwindow=true,      
     colorlinks=true,       
     linkcolor=blue,          
     citecolor=blue,        
     filecolor=magenta,      
     urlcolor=blue           
}
\usepackage[toc,page]{appendix}

\begin{document}
\title{All-magnonic Stern-Gerlach effect in antiferromagnets}
\author{Zhenyu Wang}
\affiliation{School of Electronic Science and Engineering and State Key Laboratory of Electronic Thin Films and Integrated Devices, University of Electronic Science and Technology of China, Chengdu 610054, China}
\author{Weiwei Bao}
\affiliation{School of Electronic Science and Engineering and State Key Laboratory of Electronic Thin Films and Integrated Devices, University of Electronic Science and Technology of China, Chengdu 610054, China}
\author{Yunshan Cao}
\email[Corresponding author: ]{yunshan.cao@uestc.edu.cn}
\affiliation{School of Electronic Science and Engineering and State Key Laboratory of Electronic Thin Films and Integrated Devices, University of Electronic Science and Technology of China, Chengdu 610054, China}
\author{Peng Yan}
\email[Corresponding author: ]{yan@uestc.edu.cn}
\affiliation{School of Electronic Science and Engineering and State Key Laboratory of Electronic Thin Films and Integrated Devices, University of Electronic Science and Technology of China, Chengdu 610054, China}

\date{\today}

\begin{abstract}
The Stern-Gerlach (SG) effect is well known as the spin-dependent splitting of a beam of atoms carrying magnetic moments by a magnetic-field gradient, leading to the concept of electron spin. Antiferromagnets can accommodate two magnon modes with opposite spin polarizations, which is equivalent to the spin property of electrons. Here, we propose the existence of an all-magnonic SG effect in antiferromagnetic magnonic system, where a linearly polarized spin-wave beam is deflected by a straight Dzyaloshinskii-Moriya interaction (DMI) interface into two opposite polarized spin-wave beams propagating in two discrete directions. Moreover, we observe bi-focusing of antiferromagnetic spin waves induced by a curved DMI interface, which can also spatially separate thermal magnons with opposite polarizations. Our findings provide a unique perspective to understand the rich phenomena associated with antiferromagnetic magnon spin and would be helpful for polarization-dependent application of antiferromagnetic spintronic devices.
\end{abstract}

\maketitle
\begin{figure}
  \centering
  \includegraphics[width=0.4\textwidth]{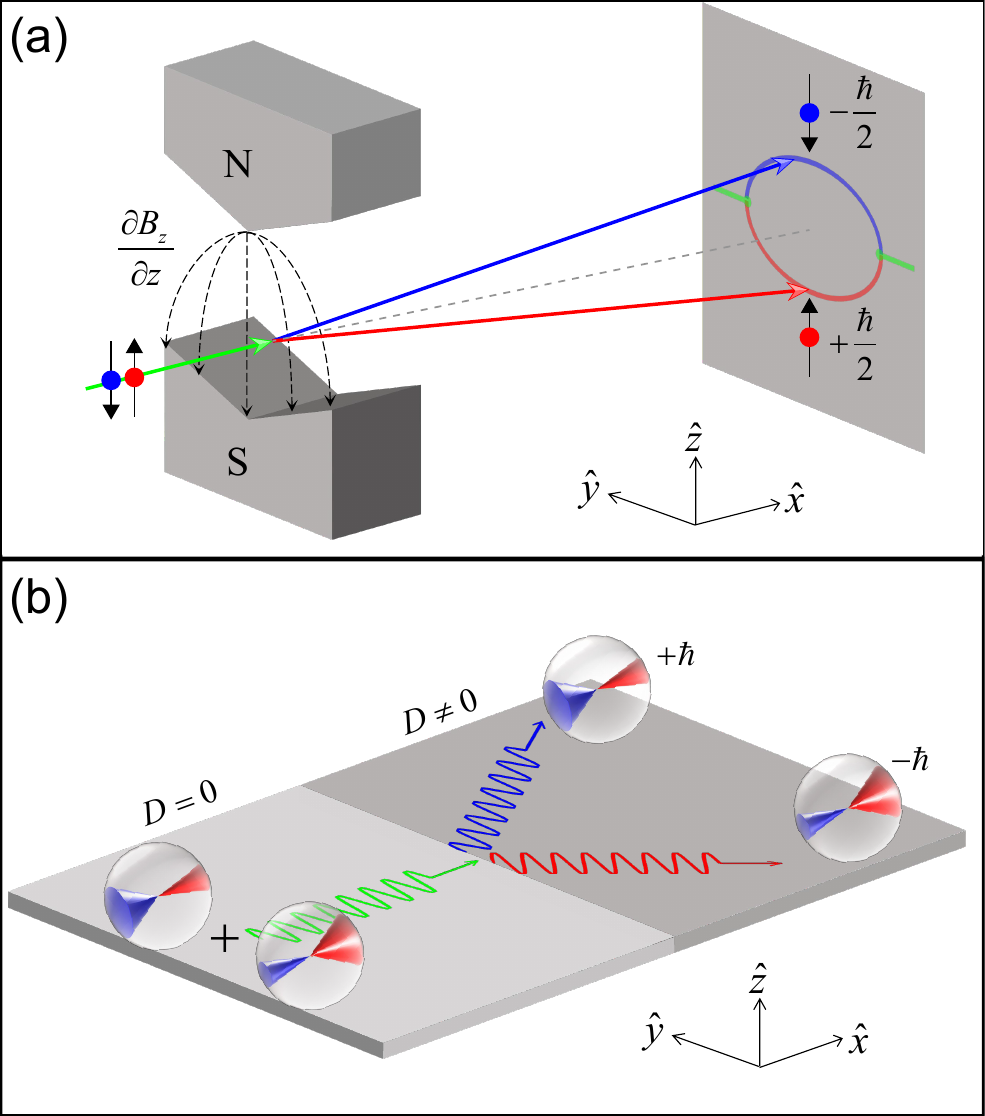}\\
  \caption{(a) Setup of the SG experiment. A beam of silver atoms, each carrying a net spin-$\frac{1}{2}$, is deflected into two discrete directions by a magnetic field gradient. (b) Schematic of the all-magnonic SG effect. A linearly polarized spin-wave beam composed of a superposition of the left- and right-handed modes is incident on a DMI region, and is deflected into two beams with opposite spin angular momenta $\pm h$.}\label{fig1}
\end{figure}

The Stern-Gerlach (SG) effect, now almost 100 years old, is one of the milestones in the development of quantum mechanics. In the original SG experiment \cite{Gerlach1922}, a beam of silver atoms passes through a region under a magnetic field gradient, and is deflected into two discrete directions, as illustrated in Fig. \ref{fig1}(a). The silver atoms only have one valence electron which contributes to the net magnetic moment, so the results reflect the properties of an electron spin. This kind of effect has been predicted in many other systems, such as photons \cite{Karpa2006,Karnieli2018,Arteaga2019}, spinor Fermi and Bose gases in tight atom waveguides \cite{Girardeau2004}, and mixed left- and right-handed chiral molecules \cite{Li2007}.

Similar to electrons, magnons also carry spin angular momentum (SAM). The SAM of a magnon is either $+\hbar$ or $-\hbar$ associated with the right- and left-handed circular polarization states of spin waves. However, in ferromagnets, only the right-handed polarized spin waves can be accommodated. For this reason, the polarization property of spin waves is rarely utilized in magnonics. Exceptions include the magnonic spin transfer torque in driving the domain wall propagation \cite{Yan2011}.
Recently, antiferromagnets with two opposite magnetic sublattices have attracted significant attention owing to their unique properties such as the ultrafast spin dynamics and vanishing stray magnetic field \cite{Jungwirth2016,Baltz2018,Jungfleisch2018}. In particular, antiferromagnet has both left- and right-handed polarized spin waves gaining the full freedom in polarization, and is regarded as a platform for magnonics superior to ferromagnet.
The coexistence of both spin polarizations in antiferromagnetic magnons \cite{Rezende2019} enables the magnonic realization of many physical phenomena associated with the electron spin, such as antiferromagnetic spin-wave field-effect transistor \cite{Cheng201601}, magnonic Nernst effect \cite{Cheng201602,Zyuzin2016}, magnonic analog of relativistic Zitterbewegung \cite{Wang2017}, and magnonic Hanle effect \cite{Wimmer2020}. Despite these analogies, however, the analogue of the SG effect in magnonic system is yet to be realized.

The Dzyaloshinskii-Moriya interaction (DMI) \cite{Dzyaloshinsky1958,Moriya1960}, present in magnetic systems with broken inversion symmetry, has a chiral character and causes the nonreciprocal propagation of spin waves, which provides additional functionalities in magnonic devices \cite{Lan2015,Yu2016,Wang2018,Wang202001}. Recent studies found that the DMI can lift the degeneracy between two polarized spin-wave modes \cite{Cheng201601,Cheng201602}, which offers the possibility of the realization of the magnonic SG effect. In this letter, we demonstrate that a DMI interface acts the equivalent of a spatially varying magnetic field, and can deflect a linearly polarized spin-wave beam into two discrete beams with opposite polarizations, as shown in Fig. \ref{fig1}(b).

Spin current, the flow of spin angular momentum, is a key concept in spintronics and can be transported by magnons in magnetic insulators.
Efficient generation of spin current is a prerequisite for practical application of antiferromagnetic spintronics. Because two polarized spin-wave modes carry opposite spin angular momenta, the net spin angular momentum of antiferromagnetic magnons is zero when they flow along the same direction. To generate spin current, the degeneracy of two polarized spin-wave modes should be broken, which is usually achieved by applying a very large magnetic field \cite{Li2020}. Although some approaches, including temperature gradients \cite{Seki2015,Wu2016} and ultrafast laser pulses \cite{Qiu2021}, have been adopted to generate spin currents in antiferromagnets, a simple yet efficient way for the spin-current generation is still lacking. The magnonic SG effect predicted in this work can lead to the spatial separation of opposite spin-wave polarizations in antiferromagnets, which offers a field-free method to generate spin currents with two spin polarizations simultaneously.

A G-type antiferromagnetic film with in-plane uniaxial anisotropy along $\hat{x}$ is considered. We define the total and staggered magnetization as follows: $\mathbf{m}\equiv(\mathbf{m}_{1}+\mathbf{m}_{2})/2$ and $\mathbf{n}\equiv(\mathbf{m}_{1}-\mathbf{m}_{2})/2$, where $\mathbf{m}_{1}$ and $\mathbf{m}_{2}$ are the reduced magnetization of two sublattices, respectively. Under the continuum approximation, the free energy density of the antiferromagnetic system is given by
\begin{equation}\label{eq_FreeEnergyDensity}
  \begin{split}
    \mathcal{U}= & \frac{\lambda}{2}\mathbf{m}^{2}+\frac{A}{2}[(\partial_{x}\mathbf{n})^{2}+(\partial_{y}\mathbf{n})^{2}+\partial_{x}\mathbf{n}\cdot\partial_{y}\mathbf{n}]\\
      & +L\mathbf{m}\cdot(\partial_{x}\mathbf{n}+\partial_{y}\mathbf{n})-\frac{K}{2}(\mathbf{n}\cdot\hat{x})^{2}+w_{D},
  \end{split}
\end{equation}
where $\lambda$, $A$, $L$, and $K$ are the homogeneous exchange, inhomogeneous exchange, parity-breaking, and magnetic anisotropy constants, respectively. $w_{D}=\frac{D}{2}[n_{z}\nabla\cdot\mathbf{n}-(\mathbf{n}\cdot\nabla)n_{z}]$ denotes the DMI energy density of the interfacial form.

With the constraint $\mathbf{n}\cdot\mathbf{m}=0$, the dynamic equation for the staggered magnetization $\mathbf{n}$ without the damping is simplified as
\begin{equation}\label{eq_d2tn}
\begin{split}
  \mathbf{n}\times\partial_{t}^{2}\mathbf{n}= & \gamma^{2}\lambda\mathbf{n}\times[A\nabla^{2}\mathbf{n}+Kn_{x}\hat{x} \\
    & +D(\hat{y}\times\partial_{x}\mathbf{n}-\hat{x}\times\partial_{y}\mathbf{n})].
\end{split}
\end{equation}
To determine the spectrum of antiferromagnetic spin waves, we assume a small fluctuation of $\mathbf{n}$ around the static staggered magnetization $\mathbf{n}_{0}=\hat{x}$, and express the staggered magnetization as $\mathbf{n}=(1,n_{y},n_{z})$ with $|n_{y,z}|\ll1$. By linearizing Eq. (\ref{eq_d2tn}) in terms of the small deviation ($n_{y},n_{z}$), and defining $n_{\pm}=n_{y}\pm in_{z}$ to describe the right- and left-handed polarized modes, we obtain a two-component Klein-Gordon equation \cite{Cheng201601}
\begin{equation}\label{eq_KG}
  \partial_{t}^{2}\Psi = \gamma^{2}\lambda[(\frac{A}{2}\nabla^{2}-K)\Psi-iD\sigma_{z}\partial_{y}\Psi],
\end{equation}
where $\Psi=(n_{+},n_{-})^{T}$ is the two-component Dirac spinor and
$\sigma_{z}=\left[
  \begin{array}{cc}
    1 & 0 \\
    0 & -1 \\
  \end{array}
\right]$ is the Pauli matrix.

\begin{figure}
  \centering
  \includegraphics[width=0.5\textwidth]{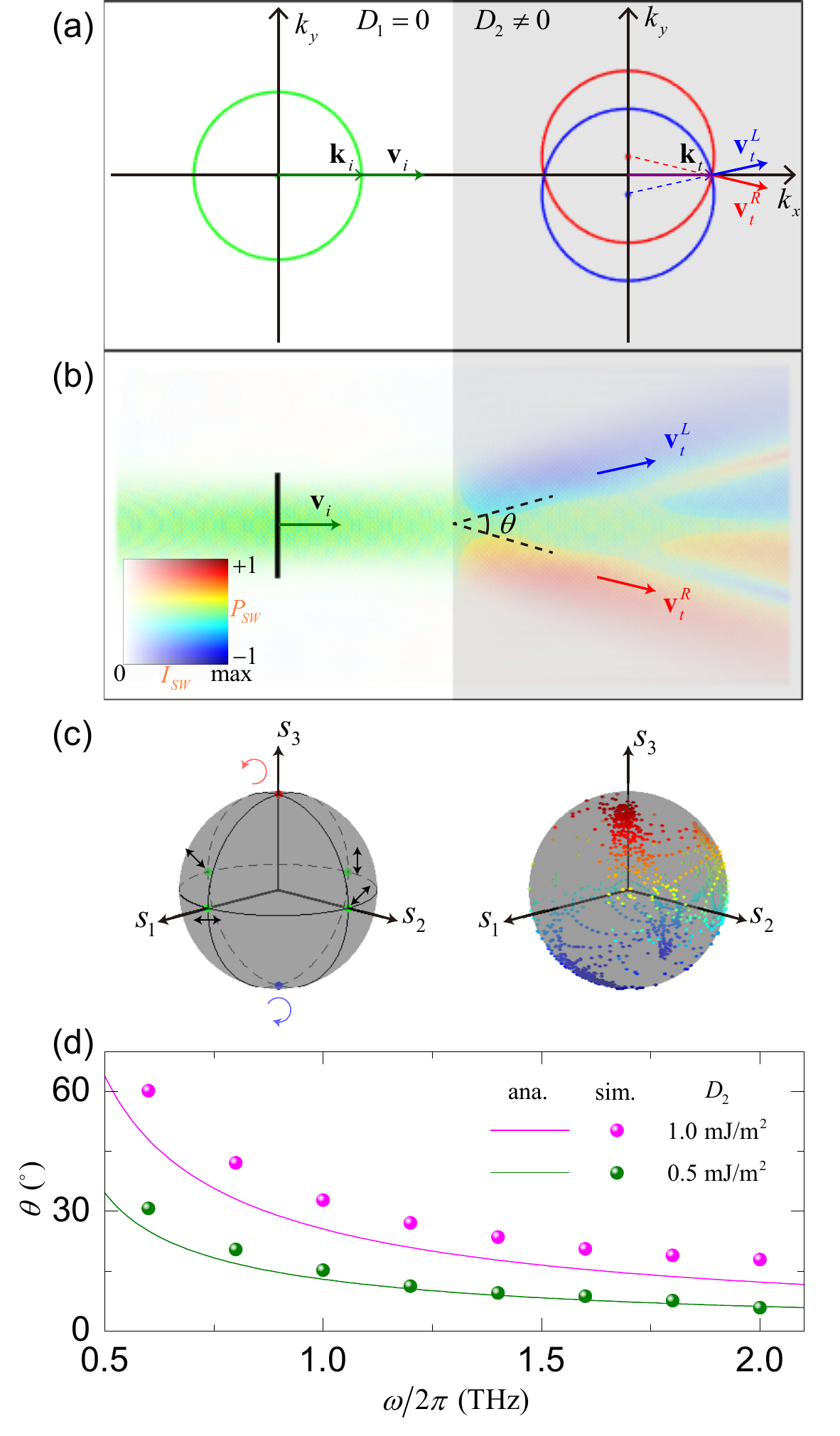}\\
  \caption{(a) The isofrequency curves of spin waves propagating in no-DMI (left) and DMI (right) regions. Blue and red circles are isofrequency curves of the left- and right-handed polarized spin waves, respectively. (b) A linearly polarized spin-wave beam with $\omega/2\pi=1$ THz propagates through the DMI interface and is divided into two spin-wave beams with opposite polarizations. The DMI constant in the right region is $D=1.0$ $\mathrm{mJ/m^{2}}$. The black bar denotes the exciting source of spin waves. The inset is a two-dimensional colorbar with color indicating the spin-wave polarization and intensity representing the spin-wave amplitude. (c) Left panel: The classical Poincar\'{e} sphere for polarization states of light wave. Right panel: The Poincar\'{e} sphere of the spin-wave polarization on the right region in (b). (d) The deflection angle $\theta$ as a function of the spin-wave frequency $\omega$ for $D_{2}=1.0$ and 0.5 $\mathrm{mJ/m^{2}}$. The dots correspond to the simulation data and the solid curves represent the analytical formula (\ref{eq_tht}).}\label{fig2}
\end{figure}

Using the plane-wave ansatz $\Psi\sim \exp[i(\mathbf{k^{\pm}}\cdot\mathbf{r}-\omega t)]$, we obtain the dispersion relation of antiferromagnetic spin waves
\begin{equation}\label{eq_dispersion}
  \omega=\gamma\sqrt{\lambda\Big[\frac{A}{2}(\mathbf{k^{\pm}})^{2}+K\mp D k_{y}^{\pm}\Big]},
\end{equation}
where the superscript $``\pm"$ corresponds to the right- and left-handed modes, respectively.
Based on Eq. (\ref{eq_dispersion}), we plot the isofrequency curve of antiferromagnetic spin waves in Fig. \ref{fig2}(a). In the DMI region, the isofrequency curves of two polarized spin waves shift oppositely along the $k_{y}$ direction, which implies that the degeneracy of two polarized modes is broken by the DMI.
The propagating direction of the spin-wave beam is determined by its group velocity
\begin{equation}\label{eq_vg}
  \mathbf{v}_{g}^{\pm}=\frac{\partial\omega}{\partial\mathbf{k^{\pm}}}=\frac{\gamma^{2}\lambda}{2\omega}(A \mathbf{k}^{\pm}\mp D\hat{y}),
\end{equation}
which indicates that two polarized spin-wave beams would be separated along the $y$ direction in the DMI region. This is schematically illustrated in Fig. \ref{fig1}(b), showing that the DMI interface splits a linearly-polarized spin-wave beam into two beams with opposite polarizations. According to Eq. (\ref{eq_vg}), the deflection angle $\theta$ between two polarized beams can be derived as
\begin{equation}\label{eq_tht}
  \cos\theta=\frac{\mathbf{v}_{g}^{+}\cdot\mathbf{v}_{g}^{-}}{|\mathbf{v}_{g}^{+}||\mathbf{v}_{g}^{-}|}
  =\frac{2A[(\omega/\gamma)^{2}/\lambda-K]-D^{2}}{2A[(\omega/\gamma)^{2}/\lambda-K]+D^{2}}.
\end{equation}

To verify our theoretical predictions, we perform full micromagnetic simulations using Mumax3 \cite{Vansteenkiste2014}. We consider a heterogeneous antiferromagnetic film with different DMI constants and adopt the following magnetic parameters \cite{Barker2016}: $A=6.59$ $\mathrm{pJ/m}$, $M_{s}=3.76\times10^{5}$ $\mathrm{A/m}$, $K=1.16\times10^{5}$ $\mathrm{J/m^{3}}$, $\lambda=150.9$ $\mathrm{MJ/m^3}$, and $D=1.0$ $\mathrm{mJ/m^{2}}$. The mesh size of $1\times1\times1$ $\mathrm{nm^{3}}$ is used to discrete the antiferromagnetic film with the size $2000\times1000\times1$ $\mathrm{nm^{3}}$. A Gilbert damping constant of $\alpha=10^{-3}$ is used to ensure a long-distance propagation of spin waves, and absorbing boundary conditions are adopted to avoid the spin-wave reflection by film edges \cite{Venkat2018}.

As shown in Fig. \ref{fig2}(b), a linearly-polarized spin-wave beam with the frequency 1 THz is excited in the left domain without the DMI, and then propagates through the DMI interface. The spatial separation of two polarized spin-wave beams is clearly observed in the right domain. This phenomenon manifests a magnonic SG effect.
The deflection angle between two polarized spin-wave beams extracted from simulation results is plotted in Fig. \ref{fig2}(d). One can see that the deflection angle obtained from simulations is a little larger than that from analytical calculations. This deviation is due to the spin canting at the DMI interface \cite{Wang2018}, and is mitigated for a small DMI step ($D_{2}=0.5$ $\mathrm{mJ/m^{2}}$), because the canting angle of the magnetization at the DMI interface decreases.
Moreover, spin canting at the DMI interface would also cause the wave-vector broadening of spin waves, leading to a special interference pattern between two beams [see Fig. \ref{fig2}(b)].

It is noted that the Poincar\'{e} sphere is a prominent graphical presentation of light's polarization \cite{Poincare1892}. In the Poincar\'{e} sphere, left- and right-handed circularly polarized states are located at two poles and linearly polarized states occupy at the equator, [see blue, red and green dots in the left panel of Fig. \ref{fig2}(c)]. Points between the equator and poles represent the intermediate elliptical polarizations. Right-handed elliptically polarized states occupy the northern hemisphere, while left-handed polarized elliptically polarized states occupy the southern hemisphere.
To determine the DMI effect on the polarization of antiferromagnetic magnons, we here map the spin-wave polarization in the right domain onto a Poincar\'{e}-like sphere through the Stokes parameters as Cartesian coordinates [see the right panel of Fig. \ref{fig2}(c)]. The Stokes parameters are determined by the amplitudes of the dynamical magnetization ($A_{y,z}$) and their phase difference ($\delta=\phi_z-\phi_y$):
\begin{subequations}
\begin{align}
&S_1=(A_y^2-A_z^2)/S_0,\\
&S_2=2A_y A_z \cos\delta/S_0,\\
&S_3=2A_y A_z \sin\delta/S_0,
\end{align}\label{eq_Stokes}\end{subequations}
where $S_0=A_y^2+A_z^2$. One can see that there exist many elliptical polarization states besides the right- and left-handed polarization states at the two poles. These elliptical states are generated by the interference effect and superposition of the left- and right-handed circularly polarized states. On the one hand, this phenomenon could be mitigated by improving the collimation of the linearly polarized spin-wave beam. On the other hand, it provides an appealing approach to generate arbitrary spin-wave polarizations in insulating antiferromagnets.

\begin{figure}
  \centering
  \includegraphics[width=0.5\textwidth]{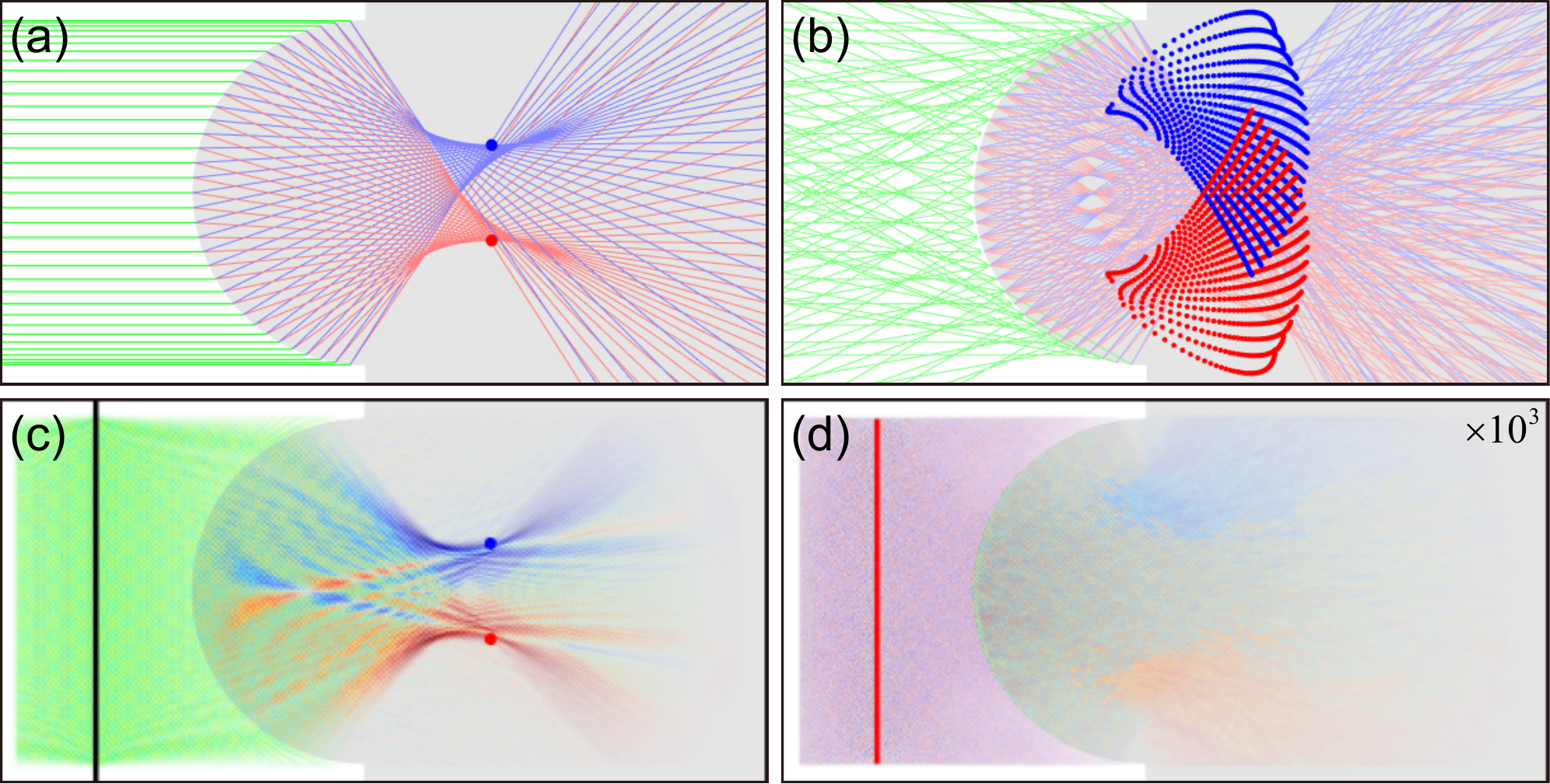}\\
  \caption{Coherent [(a) and (c)] and incoherent [(b) and (d)] spin waves propagation through the interface between two antiferromagnetic films with different exchange constants ($A_{2}=0.5A_{1}$) and DMIs ($D_{1}=0$, $D_{2}=0.7$ $\mathrm{mJ/m^{2}}$). (a) and (b) are obtained by the analytical formula (\ref{eq_fp}). Green, blue, and red lines represent the incident linearly polarized, transmitted left- and right-handed spin-wave beams. (c) and (d) are numerical simulations. The black bar in (c) denotes the exciting source of coherent spin waves with 1 THz. The red bar in (d) denotes the thermal source at temperature 100 K. The spin-wave intensity of the right part in (d) is magnified by a factor of $10^{3}$ for ease of observation. Blue and red dots correspond to the focal points of the left- and right-handed spin waves, respectively.
  }\label{fig3}
\end{figure}

In ferromagnets, it has been demonstrated that a semicircular DMI interface can cause the off-axis focusing of spin waves \cite{Bao2020}. This phenomenon can be extended naturally to antiferromagnetic spin waves, with features being significantly enriched by the additional polarization degree of freedom. To focus antiferromagnetic spin waves, we design a spin-wave lens using a semicircular interface between two antiferromagnetic films with different exchange constants ($A_{1,2}$) and DMI strengths ($D_{1,2}$). Different exchange constants can be archived by doping or ion bombardment \cite{Sorensen2019,Macedo2022}, and the DMI strengths can be locally altered by using lithographic techniques to change the covering heavy-metal layers \cite{Torrejon2014,Tacchi2017}.
The propagation of the linearly polarized spin waves through a heterochiral curved interface is derived theoretically (see the supplementary material for analytical details), as illustrated in Fig. \ref{fig3}(a). One can see that left- and right-handed polarized spin waves are focused oppositely along $y$ direction, because they experience opposite effective fields induced by the DMI [based on Eq. (\ref{eq_KG})]. The corresponding focal-point coordinations are also given as (see the supplementary material)
\begin{eqnarray}\label{eq_fp}
\begin{aligned}
  x_{f}^{\pm} &=R\frac{k_{r}^{2}}{k_{r}^{2}-k_{r}^{1}}\cos\theta_{0}^{\pm}\cos^{2}(\beta+\theta_{0}^{\pm})+R(1-\cos\theta_{0}^{\pm}),
  \\
  y_{f}^{\pm} &=R\frac{k_{r}^{2}}{k_{r}^{2}-k_{r}^{1}}\tan\beta\cos\theta_{0}^{\pm}\cos^{2}(\beta+\theta_{0}^{\pm})+R\sin\theta_{0}^{\pm},
\end{aligned}
\end{eqnarray}
where $R$ is the radius of the semicircular interface, $k_{r}^{1}$ and $k_{r}^{2}$ are the radiuses of the isofrequency circles in no-DMI and DMI regions, $\beta$ is the incident angle with respect to the lens axis (along $x$ axis), and $\theta_{0}^{\pm}$ is the center angle of the incident point where the incident and refracted beams are parallel. Numerical simulation shows a good agreement with the theoretical result, as plotted in Fig. \ref{fig3}(c). We call such a phenomenon the bi-focusing effect of antiferromagnetic spin waves, which can be utilized to enhance the signal strength of the magnonic spin current in antiferromagnets.

However, the coherent excitation of polarized spin waves in antiferromagnets requires extremely high frequency ($\sim$  Terahertz) wave sources and very strong magnetic fields ($\sim$ a few Tesla) to break the degeneracy of the two polarizations, which is difficult to be experimentally realized \cite{Li2020}. Whereas incoherent spin waves can be easily generated by thermal agitations \cite{Seki2015,Zhang2020}. Thus, spatial separation of thermally excited magnons is crucially important for the spin-current generation and practical applications of antiferromagnetic magnons. Thermal magnons consist of many spin-wave states with different frequencies and propagating directions. It is expected that the above heterochiral curved interface is also able to split two polarized incoherent spin waves. To verify this feature, we calculate the focal-point coordinations of thermal spin waves propagating through such a heterochiral curved interface using Eq. (\ref{eq_fp}) and Eq. (S11) in the supplementary material [shown in Fig. \ref{fig3}(b)], which is then compared with simulation results shown in Fig. \ref{fig3}(d). We find that the polarizations of the transmitted spin waves are indeed spatially separated, suggesting that the generation of magnonic spin current using bi-focusing of spin waves is possible even for incoherent thermal magnons.

\begin{figure}
  \centering
  \includegraphics[width=0.4\textwidth]{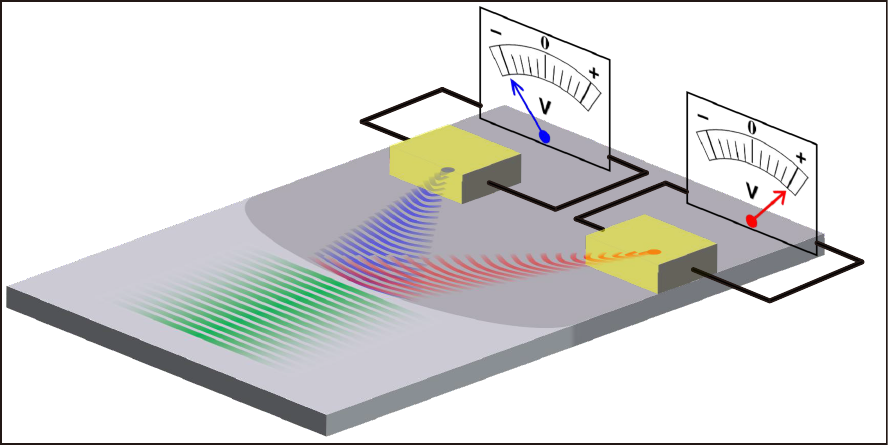}\\
  \caption{Schematic of the spin-current generation and detection.}\label{fig4}
\end{figure}

To experimentally test the spin-current generation by bi-focusing of spin waves, we propose a potential setup consisting of two heavy metals and an antiferromagnetic layer, as illustrated in Fig. \ref{fig4}. The two heavy metals locate at the focal points of the antiferromagnetic films. The focused spin waves can pump spin current into the heavy metal. Through the inverse spin Hall effect (ISHE), the spin current is converted into a DC voltage that can be monitored by the voltmeter \cite{Kimura2007,Kajiwara2010}. Because of the opposite polarizations of spin waves at two focal points, we expect that the sign of the two measured ISHE voltages would be opposite.

In summary, we predicted a magnonic SG effect by investigating the propagation of a linearly polarized spin-wave beam through a straight DMI interface. By designing a spin-wave lens via a semicircular interface between two regions with different exchange constants and DMIs, we observed a bi-focusing of antiferromagnetic spin waves, depending on their polarizations. This semicircular interface is demonstrated to be able to split the thermally excited incoherent magnons, too. These results open the door to generating spin current with opposite polarizations without applying external fields. Our findings deepen the understanding of the DMI-induced rich phenomena of antiferromagnetic magnons and are helpful for manipulating spin-wave polarization in magnonic devices.
\\

See the supplementary material for calculation details.
\\

We thank X.S. Wang, Z.-X. Li, Z. Zhang, H. Yang and C. Xiao for helpful discussions. This work was supported by the National Natural Science Foundation of China (Grants No. 12074057, 11604041, and 11704060). Z.W. acknowledges the financial support from the China Postdoctoral Science Foundation under Grant No. 2019M653063.

\end{document}